\documentclass[fleqn,twoside]{article}
\usepackage{espcrc2}

\usepackage{graphicx}
\usepackage{epsfig}

\newcommand{\AmS}{{\protect\the\textfont2
  A\kern-.1667em\lower.5ex\hbox{M}\kern-.125emS}}

\title{Non-linear Brane Dynamics in Six Dimensions}

\author{B. Cuadros-Melgar\address[IFUSP]{Instituto de F\'\i sica,
  Universidade de S\~ao Paulo \\ C.P.66.318, CEP 05315-970, S\~ao
  Paulo, Brazil}%
        \thanks{I would like to thank Elcio Abdalla for useful
discussions and for reading the manuscript. This work has been
supported by Funda\c c\~ao de Amparo \`a Pesquisa do Estado de S\~ao
Paulo {\bf (FAPESP)}, Brazil.}}
       
\begin{document}

\begin{abstract}
We consider a dynamical brane world in a six dimensional spacetime
containing a singularity. Using the Israel conditions we study the
motion of a 4-brane embedded in this setup. We analize the brane
behavior when its position is perturbed about a fixed point and solve
the full non-linear  
dynamics in the several possible scenarios. We also investigate
the possible gravitational shortcuts and
calculate the delay between graviton and photon signals
and the ratio of the corresponding subtended horizons.
\vspace{1pc}
\end{abstract}

\maketitle

\section{The Brane Cosmological Model}

We consider a six-dimensional model described by the following metric 
\begin{eqnarray}\label{metric}
ds^2 &=& -n^2(t,y,z) dt^2 + a^2(t,y,z) d\Sigma_{k} ^2 +  \nonumber \\
&& + b^2(t,y,z) dy^2 + c^2(t,y,z) dz^2 \, ,
\end{eqnarray}
where $d\Sigma_{k} ^2$ represents the metric of the three dimensional
spatial sections with $k=-1,\,0,\,1$ corresponding to a hyperbolic, a
flat and an elliptic space, respectively.

The matter content on the brane is directly related to the jump of the
extrinsic curvature tensor across the brane \cite{chr,radion}. This
relation has been derived in the case of a static brane in a previous work
\cite{accm}. Here we generalize our result 
for the Israel conditions to include the case of a brane moving with
respect to the coordinate system, which position at any bulk time $t$ is
denoted by $z = {\cal R}(t)$.

The extrinsic curvature tensor on the brane is given by
\begin{equation}\label{excurv}
K_{MN} = \eta^L _M \bigtriangledown _L \tilde n_N \, ,
\end{equation}
where $\tilde n^A$ is a unit vector field normal to the brane
worldsheet 
\begin{equation}\label{ntilde}
\tilde n^A = \left\{ {{c \, \dot{\cal R}}\over {n^2
\sqrt{1-{{c^2}\over{n^2}} \dot{\cal R}^2}}},0,0,0,0,{1 \over
{c\sqrt{1-{{c^2}\over{n^2}} \dot{\cal R}^2 }}} \right\} \,, 
\end{equation}
and
\begin{equation}\label{eta}
\eta_{MN} = g_{MN} - \tilde n_M \tilde n_N
\end{equation}
is the induced metric on the brane, from which we can obtain a
relation between $dt$ (the bulk time) and $d\tau$ (the brane time),
\begin{equation}\label{dtdtau}
d\tau = n(t,{\cal R}(t)) \sqrt{ 1- {{c^2(t,{\cal R}(t))}\over
{n^2(t,{\cal R}(t))} } \dot {\cal R}^2}\, dt \equiv n \gamma^{-1} dt \, ,
\end{equation}
where a dot means derivative with respect to the bulk time $t$. 

\subsection{The Israel Conditions}

The energy-momentum tensor on the brane located at $z_0$ can be
written as
\begin{equation}\label{em}
T^{(b)} _{MN} = {{\delta (z-z_0)} \over c} \left\{ (\rho +p) u_M u_N +
p \, \eta_{MN} \right\}\, .
\end{equation}
We also define a tensor $\hat T_{AB}$ as
\begin{equation}\label{tab}
\hat T_{AB} \equiv T_{AB} - {1 \over 4} T \eta_{AB} \, .
\end{equation}

The Israel junction conditions \cite{wisrael} are given by
\begin{equation}\label{israel}
[ K_{\mu\nu} ] = -\kappa_{(6)} ^2 \hat T_{\mu\nu} \, ,
\end{equation}
where the brackets stand for the jump across the brane and
$K_{\mu\nu}=e_\mu ^A e_\nu ^B K_{AB}$, where $e_\mu ^A$ form a basis
of the vector space tangent to the brane worldvolume.
The left-hand side of (\ref{israel}) can be calculated taking into
account the mirror symmetry across the brane.

At this point it is convenient to choose a specific bulk metric
of the form (\ref{metric}) satisfying six dimensional Einstein
equations. This is given by
\begin{equation}\label{bhmetric}
ds^2 = - h(z) dt^2 + a^2(z) d\Sigma_k ^2 + h^{-1}
(z) dz^2 \, ,
\end{equation}
where
\begin{eqnarray}
a(z) &=& {z \over l} \, , \label{az} \\
d\Sigma_k ^2 &=& {{dr^2} \over {1-kr^2}} + r^2 d\Omega_{(2)} ^2 +
(1-kr^2) dy^2 \, , \label{spacediff} \\
h(z) &=& k + {z^2 \over l^2} - {M \over z^3} + {Q^2 \over
z^6} \label{h}  
\end{eqnarray}
with $l^{-2} \propto -\Lambda$ ($\Lambda$ being the cosmological
constant) and $M$ and $Q^2$ are
constants. We should notice that $y$ is a compactified coordinate.

This metric contains a singularity located at $z=0$ and it
is valid on the $z <{\cal R}(t)$ parts of surfaces of constant
$t$, and its reflection, by the $Z_2$ orbifold symmetry, is valid on the
$z >{\cal R}(t)$ parts. If $M=0$ and 
$Q^2=0$, then (\ref{bhmetric}) is simply the metric of de Sitter or Anti
de Sitter spacetime according to the sign of $l^2$. 

With this Ansatz the Israel conditions (\ref{israel}) reduce to only
two equations, which read
\begin{eqnarray}\label{brem}
\ddot{\cal R} + {1\over 2} {{h'}\over h^3} \dot{\cal R}^4 - 3
{{h'}\over h} \dot{\cal R}^2 + {1 \over 2} h \, h' = \nonumber \\
-\kappa_{(6)} ^2
\left( {{3\rho + 4 \, p}\over 8} \right) h^2 &\left( 1- {{\dot{\cal
        R}^2} \over h^2} \right)^{3/2}& \nonumber \\ \\
{{a'} \over a} + {{\dot{\cal R}}\over h^2} {{\dot a}\over a} =
\kappa_{(6)} ^2 {\rho\over 8} \left( 1- {{\dot{\cal
        R}^2} \over h^2} \right)^{1/2} \, , \nonumber
\end{eqnarray}
where all the metric coefficients must be evaluated on the
brane. The system (\ref{brem}) describes the full non-linear dynamics
of the brane embedded in the static bulk (\ref{bhmetric}).

\subsection{The Geodesic Equation and the Time Delay}

We consider two points on the brane $r_A$ and $r_B$. In general there
are more than one null geodesics connecting them in the $1+5$
spacetime. The trajectories of photons must be on the brane and those
of gravitons may be outside. The graviton path is defined equating
(\ref{bhmetric}) to zero.
Since we are looking for a path that minimizes $t$ when the final
point $r_B$ is on the brane, the problem reduces to an
Euler-Lagrange problem \cite{acmfw}.
Then as in \cite{acm} the shortest graviton path is given by
\begin{equation}\label{geo}
\ddot {\cal R}_g + \left( {1 \over {\cal R}_g} - {3 \over 2}
  {{h'}\over h}\right) \dot {\cal R}_g^2 + {1\over 2} h \, h' - {h^2
  \over {\cal R}_g} = 0 \; . 
\end{equation}

The time delay of the photon traveling on
the brane with respect to the gravitons traveling in the bulk measured
by an observer on the brane can 
approximately be written as \cite{acm}
\begin{eqnarray}\label{delay}
\Delta\tau &\simeq& {\cal R}(t_f) \int_0 ^{t_f} dt \left({{1}\over
{{\cal R}_g(t)}} 
\sqrt{h({\cal R}_g)- {{\dot {\cal R}_g(t)^2}\over {h({\cal R}_g)}}} -
\right. \nonumber \\
&& -\left. {1 \over {{\cal R}(t)}} 
{{d\tau}\over{dt}} \right) \, .
\end{eqnarray}

It is also interesting to look at the ratio between the horizons
subtended by the photons traveling on the brane and the gravitons
traveling in the bulk,
\begin{equation}\label{ratio}
{g \over \gamma} = { {\int_0 ^{t_f} {{dt}\over {{\cal R}_g(t)}}
\sqrt{h({\cal R}_g)- {{\dot {\cal R}_g(t)^2}\over {h({\cal R}_g)}}}}
\over {\int_0 ^{t_f} {{dt}\over {{\cal R}(t)}} {{d\tau}\over{dt}}} }
\, . 
\end{equation}

\begin{figure*}[htb!]
\begin{center}
\leavevmode
\begin{eqnarray}
\epsfxsize= 3.7truecm\rotatebox{-90}
{\epsfbox{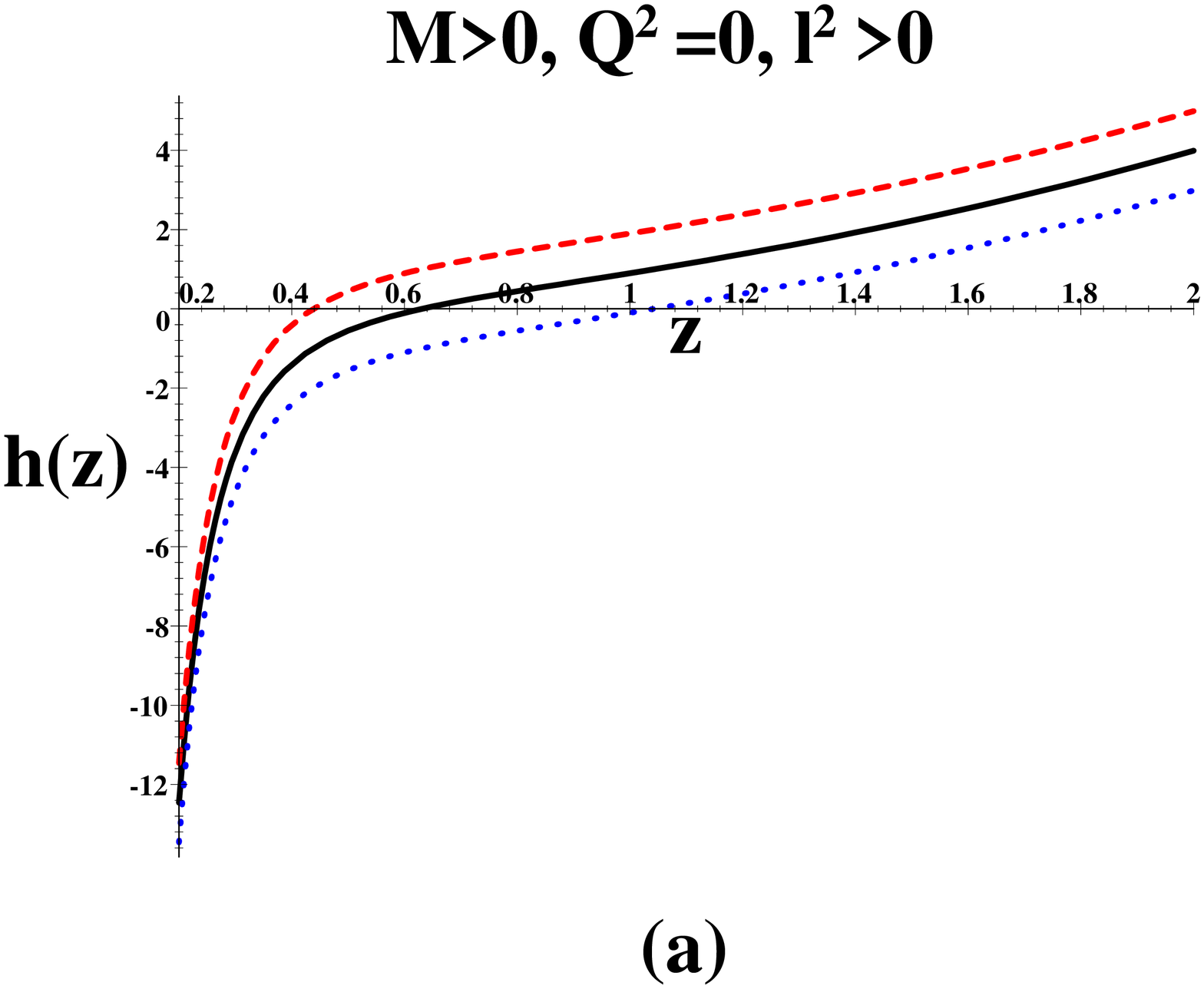}}\nonumber
\epsfxsize= 3.7truecm\rotatebox{-90}
{\epsfbox{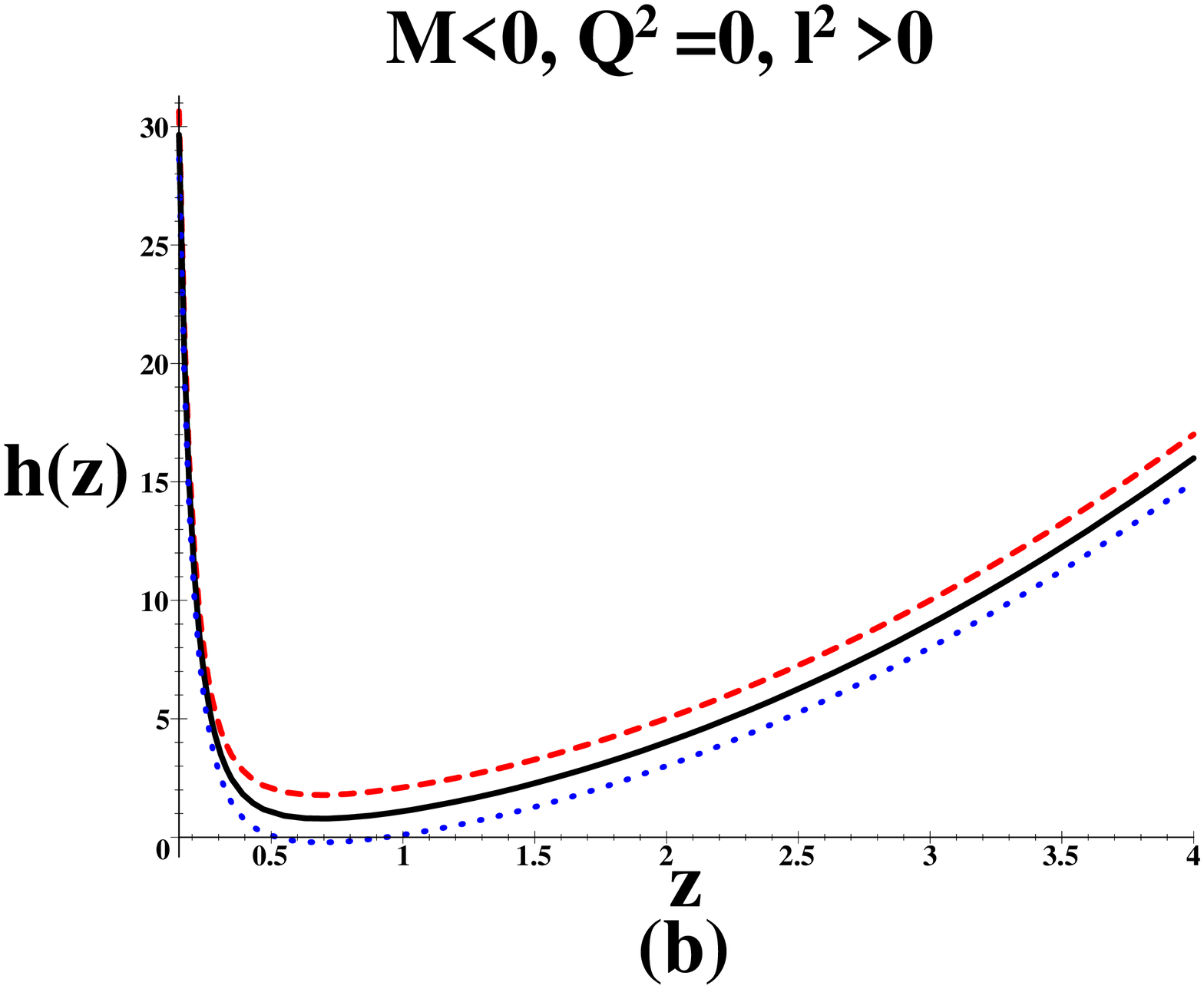}}\nonumber
\epsfxsize= 3.7truecm\rotatebox{-90}
{\epsfbox{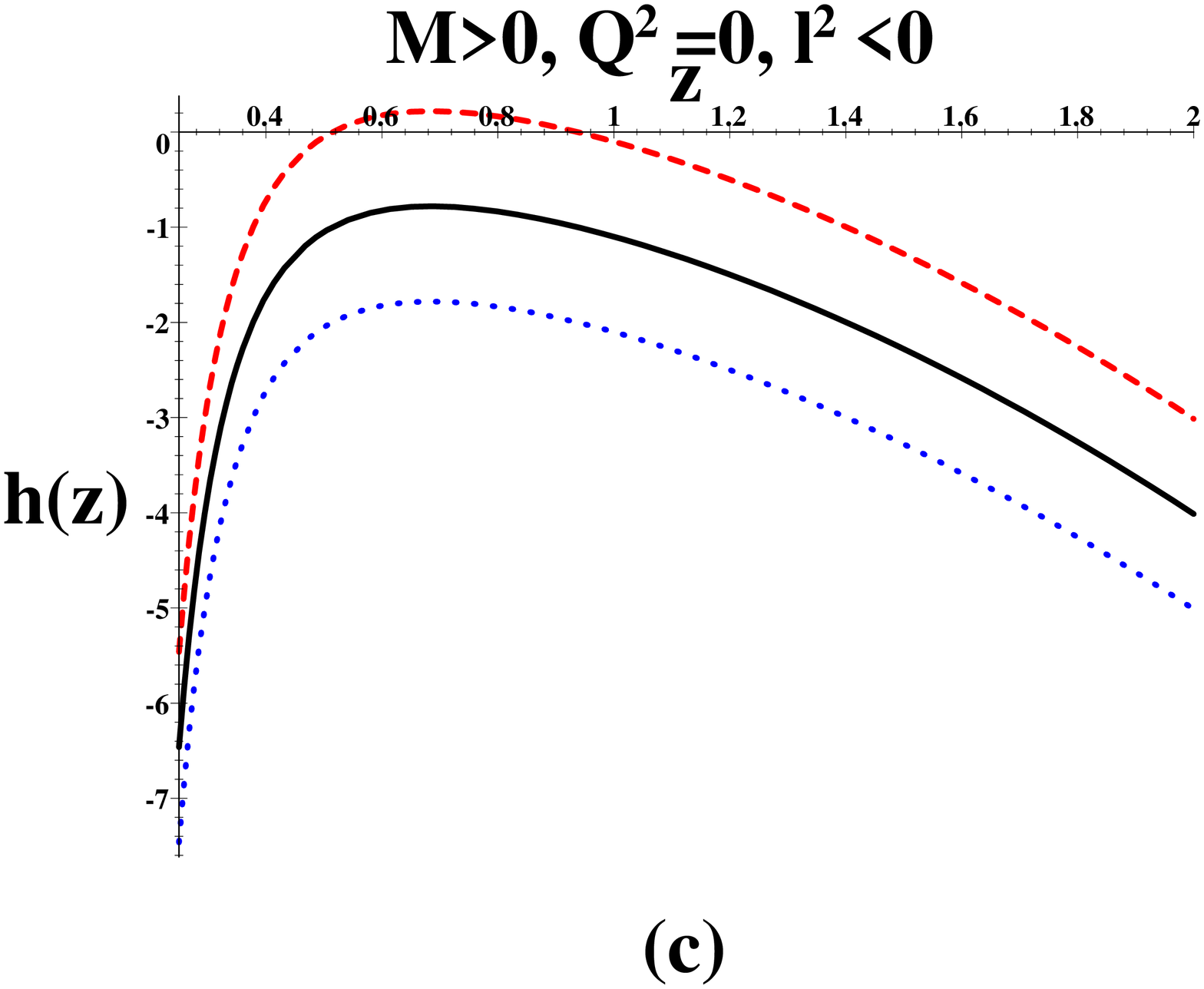}}\nonumber\\
\epsfxsize= 3.7truecm\rotatebox{-90}
{\epsfbox{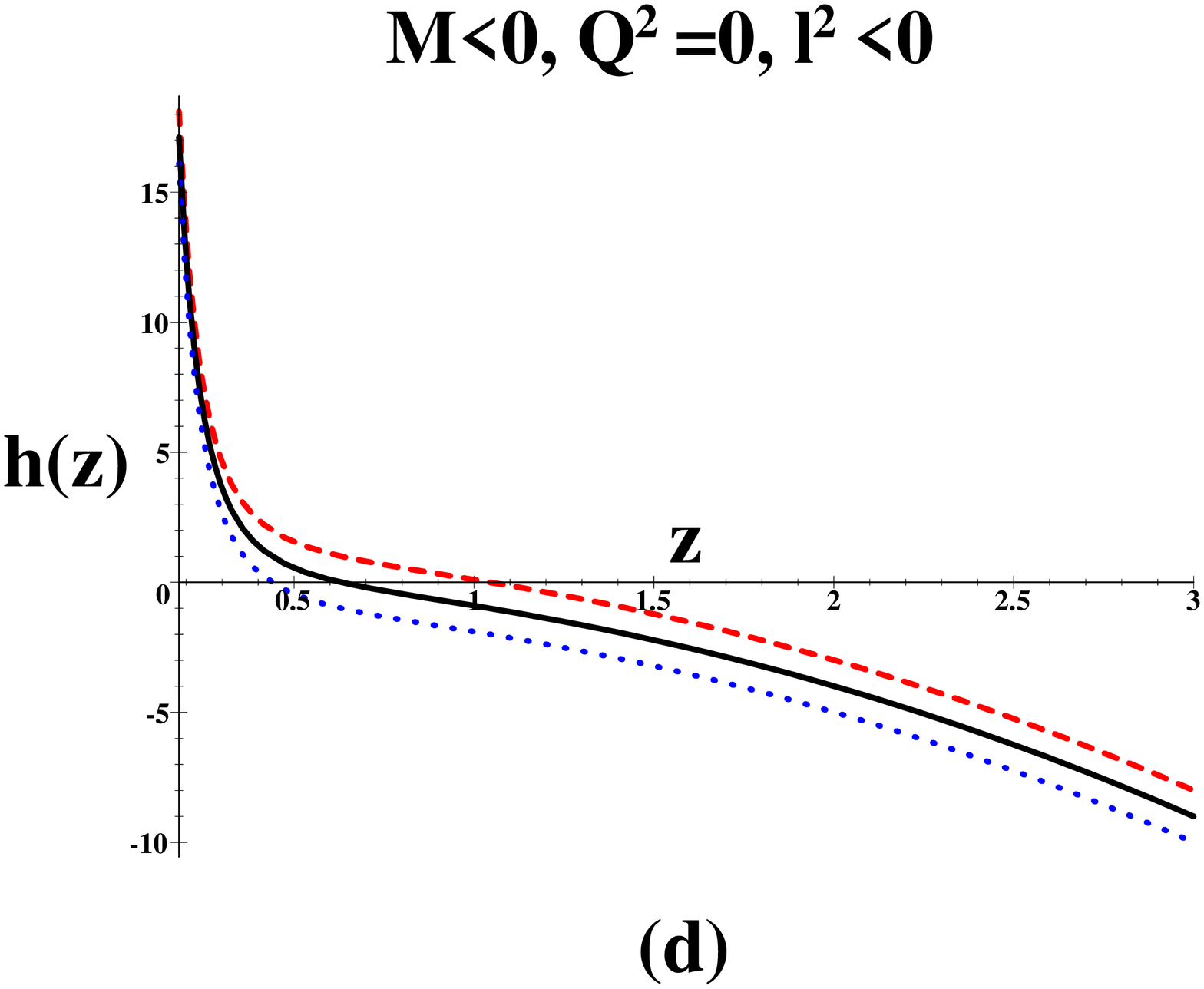}}\nonumber
\epsfxsize= 3.7truecm\rotatebox{-90}
{\epsfbox{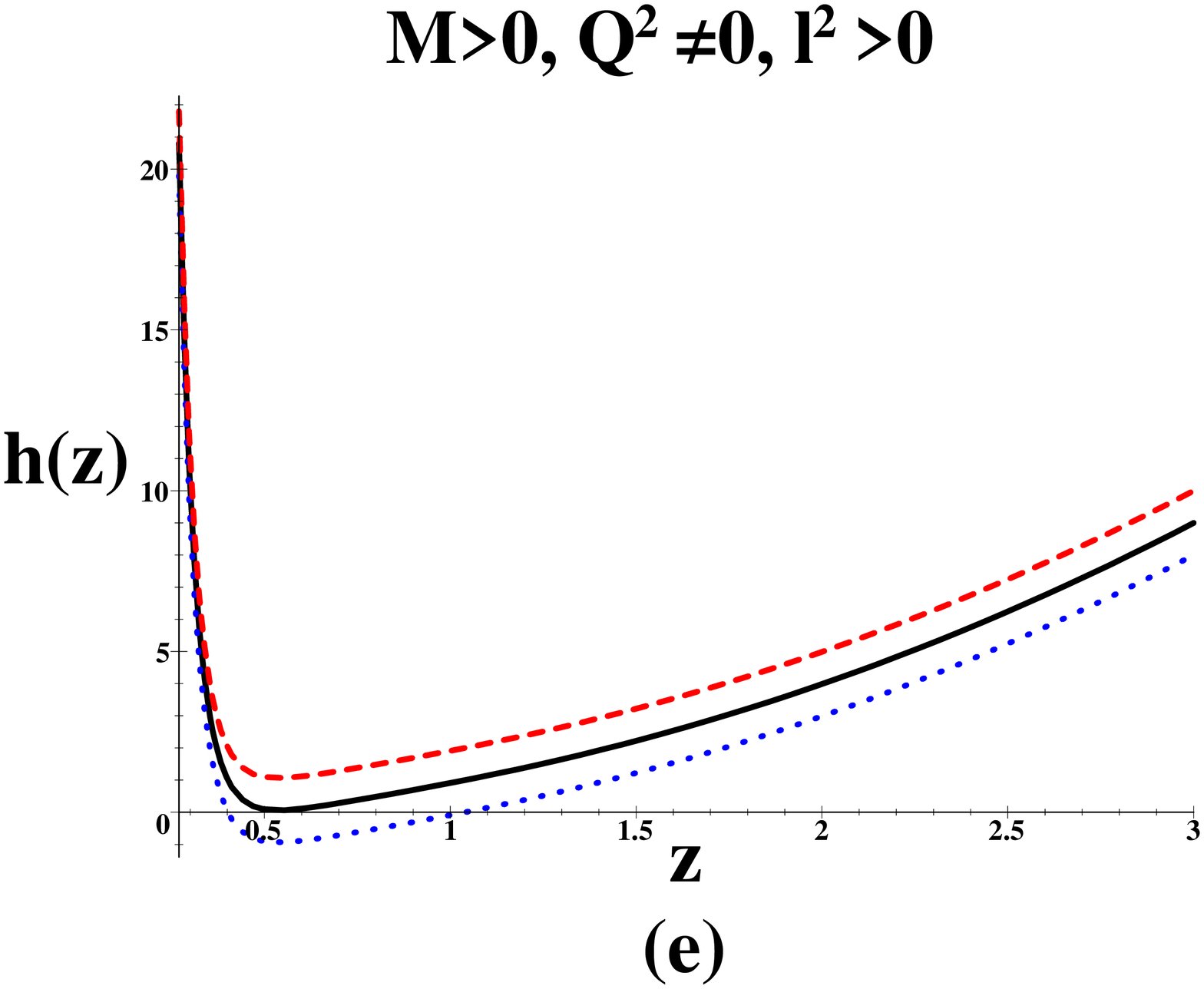}}\nonumber
\epsfxsize= 3.7truecm\rotatebox{-90}
{\epsfbox{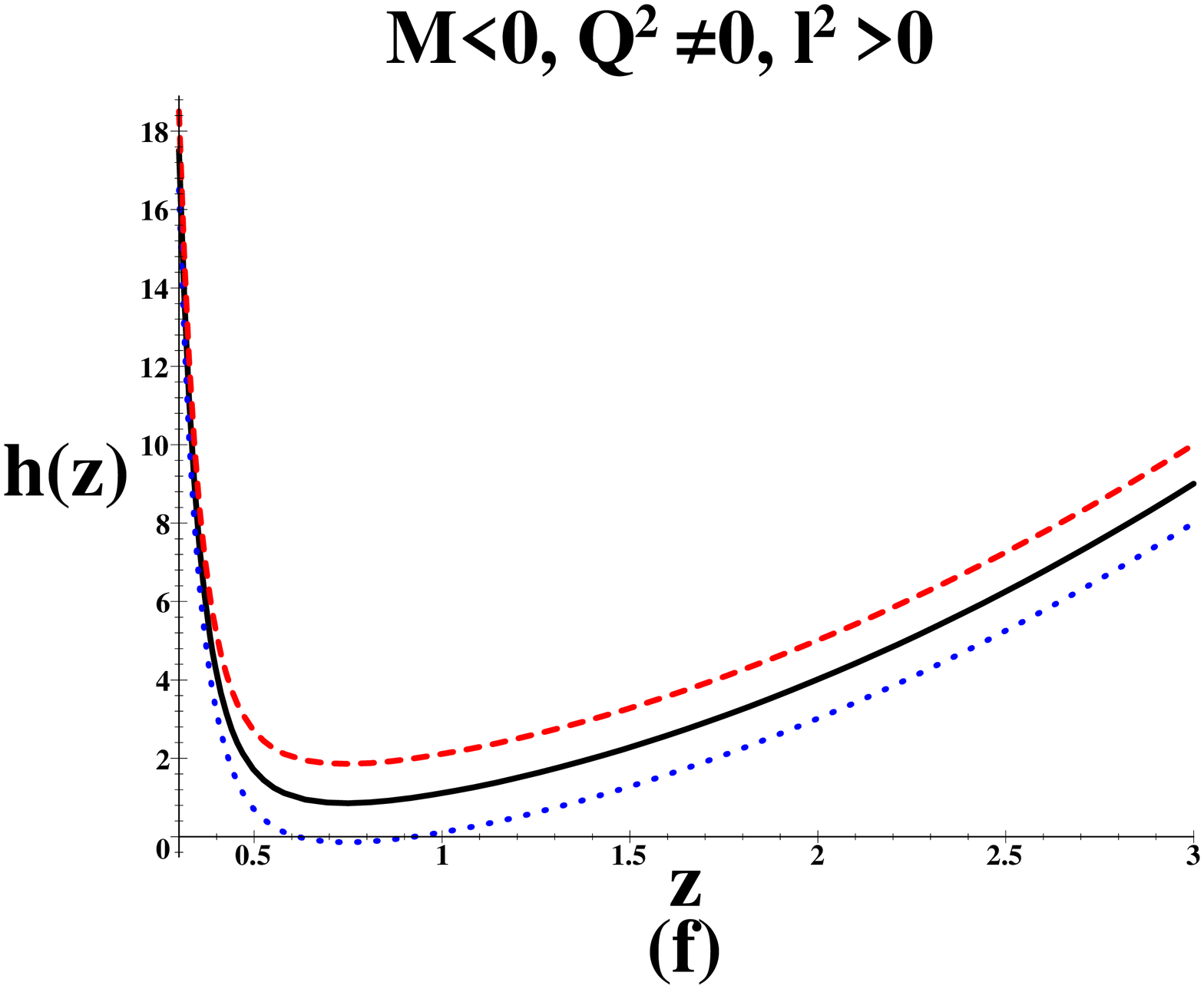}}\nonumber\\
\epsfxsize= 3.7truecm\rotatebox{-90}
{\epsfbox{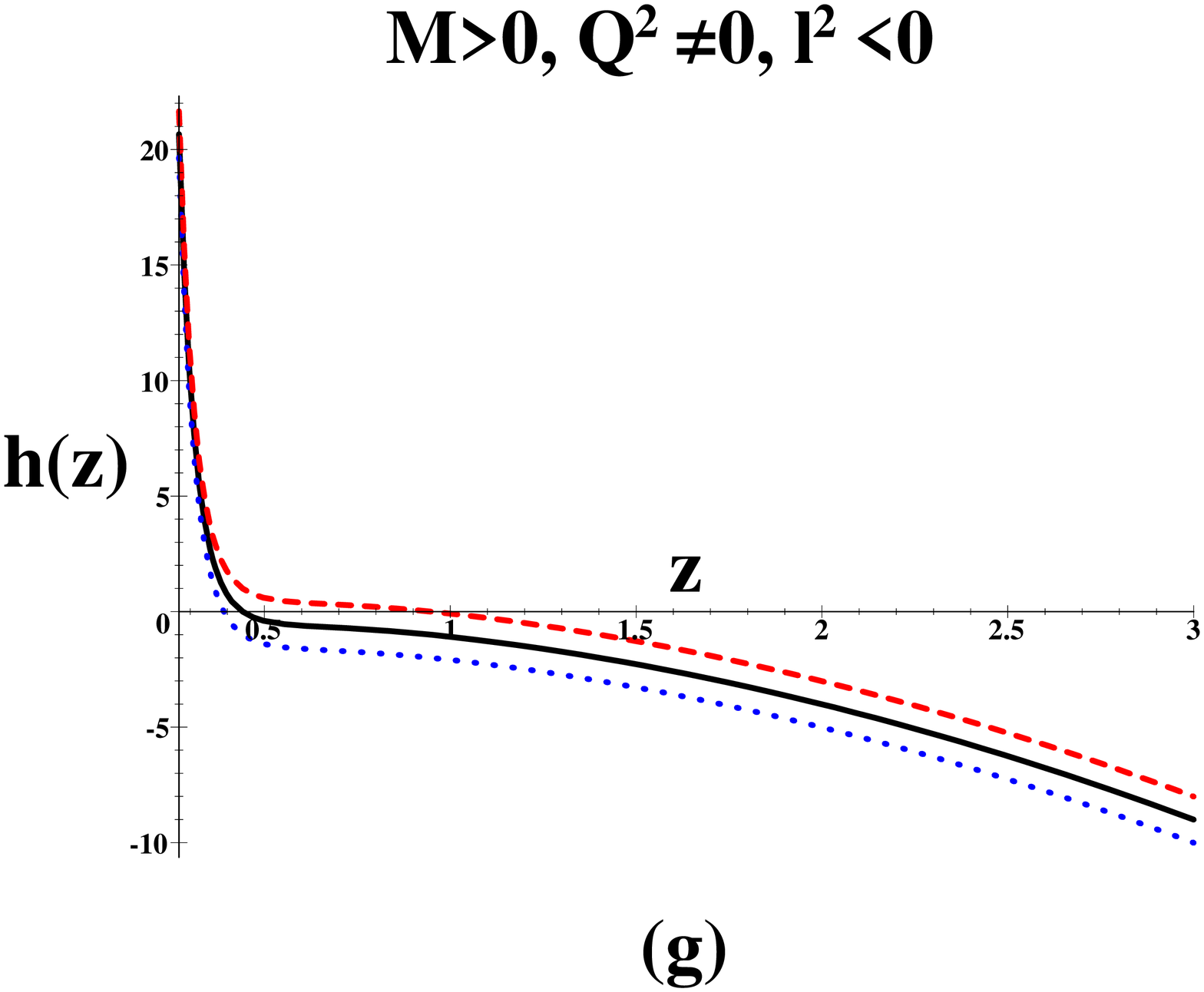}}\nonumber
\epsfxsize= 3.7truecm\rotatebox{-90}
{\epsfbox{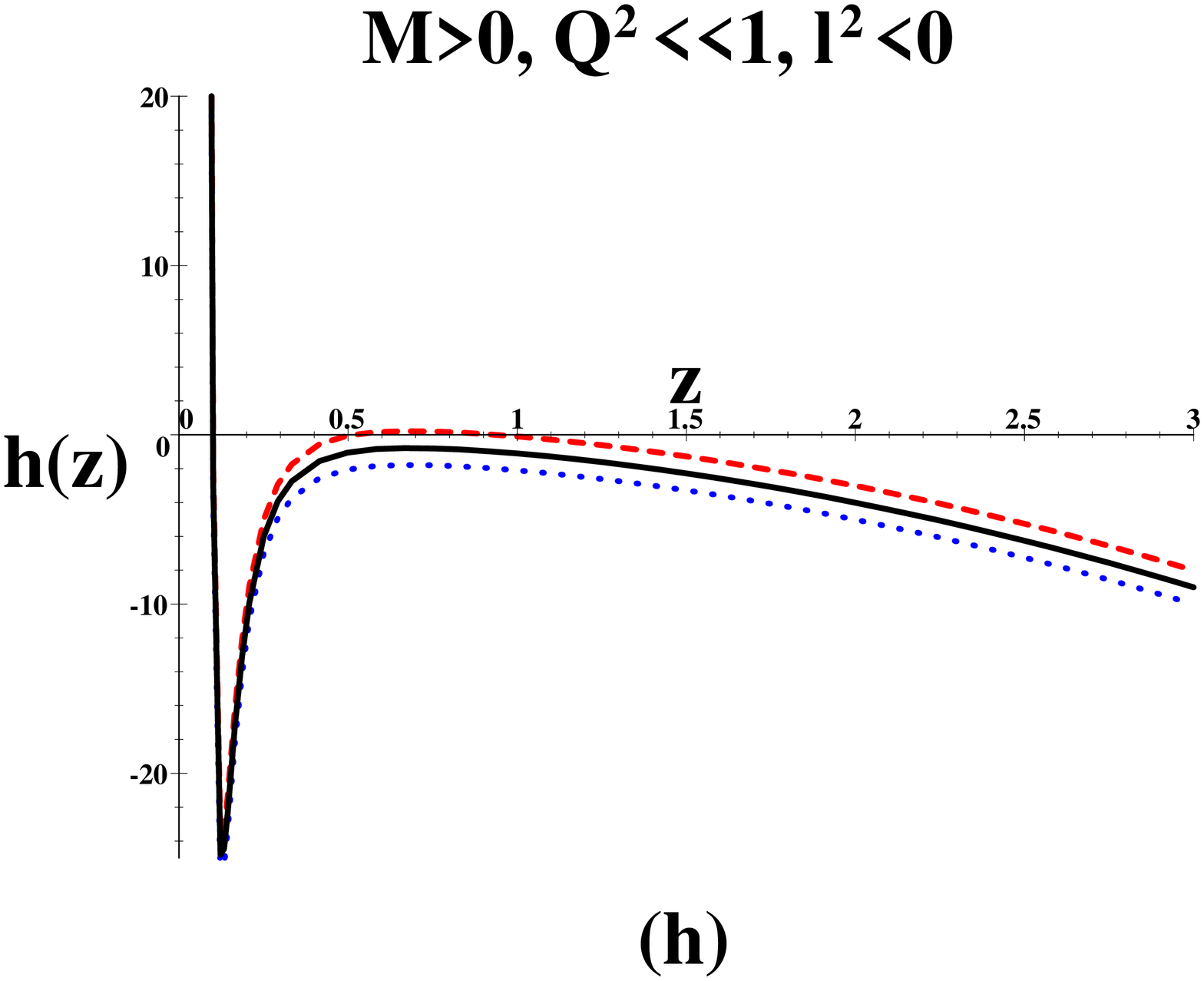}}\nonumber
\epsfxsize= 3.7truecm\rotatebox{-90}
{\epsfbox{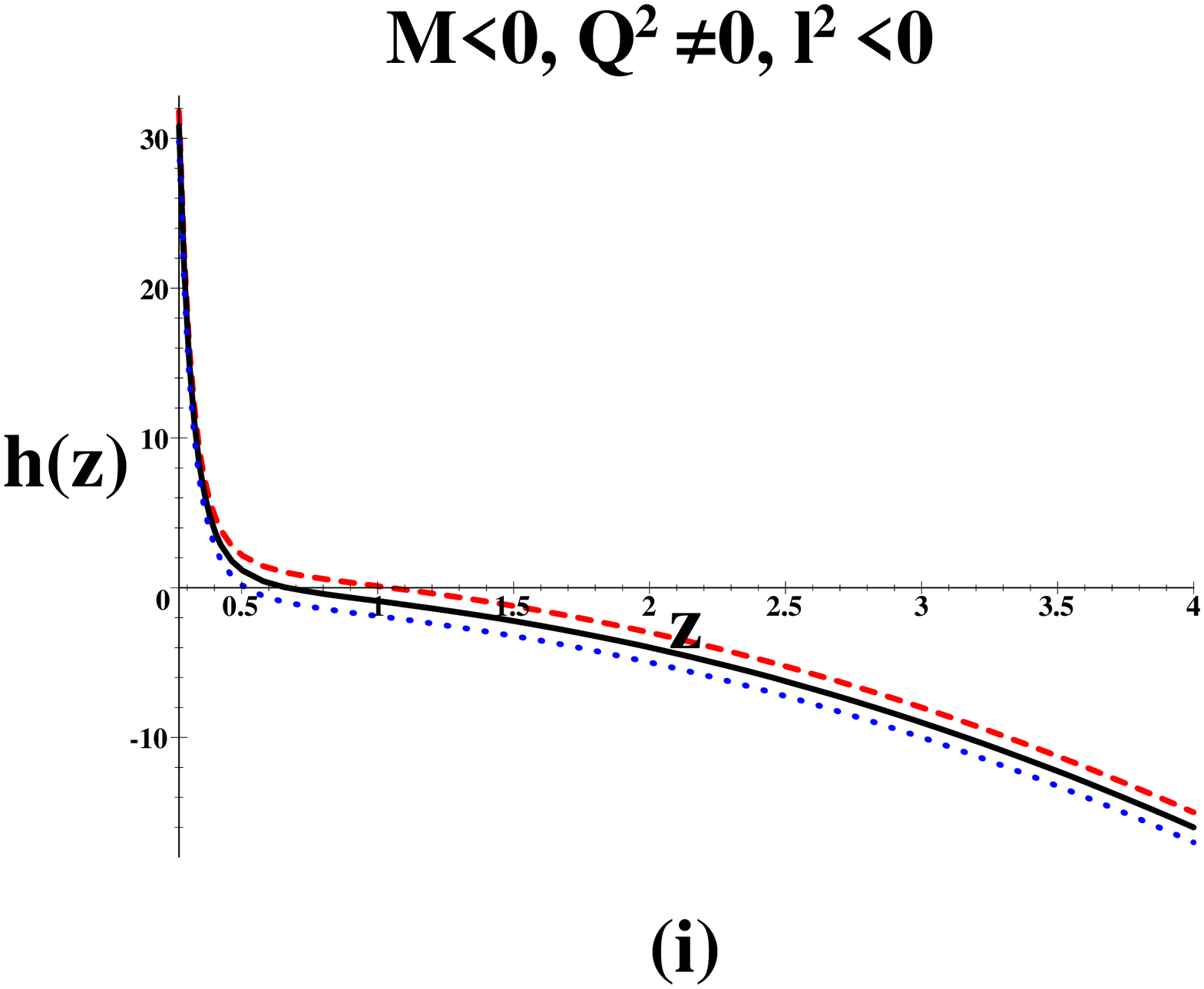}}\nonumber
\end{eqnarray}
\caption{$h(z)$ for all the possible combinations of $M$, $Q^2$ and
  $l^2$, including $k=1$ (dashed lines), $k=0$ (solid lines) and $k=-1$
  (dotted lines) cases.} 
\label{hh}
\end{center}
\end{figure*}

\section{Non-linear Brane Dynamics}

The system (\ref{brem}) describing the brane dynamics can be
numerically solved for several combinations of $M$, $Q^2$, $k$ and
$l^2$. When $M$, $Q^2$ and $k$ vanish, the solution for ${\cal R}(t)$
is a constant or a linear function in $t$ depending on the given initial
condition for $\dot{\cal R}$.

\begin{figure*}[htb!]
\caption{\label{fig6}Scale factor evolution for domain wall, matter
  and radiation 
  dominated branes and geodesics when $M>0$, $Q^2 \not= 0$ and $l^2>0$
  (Anti de Sitter - naked singularity).}
\begin{center}
\begin{eqnarray}
\rotatebox{0}
{\includegraphics[width=.50\textwidth]{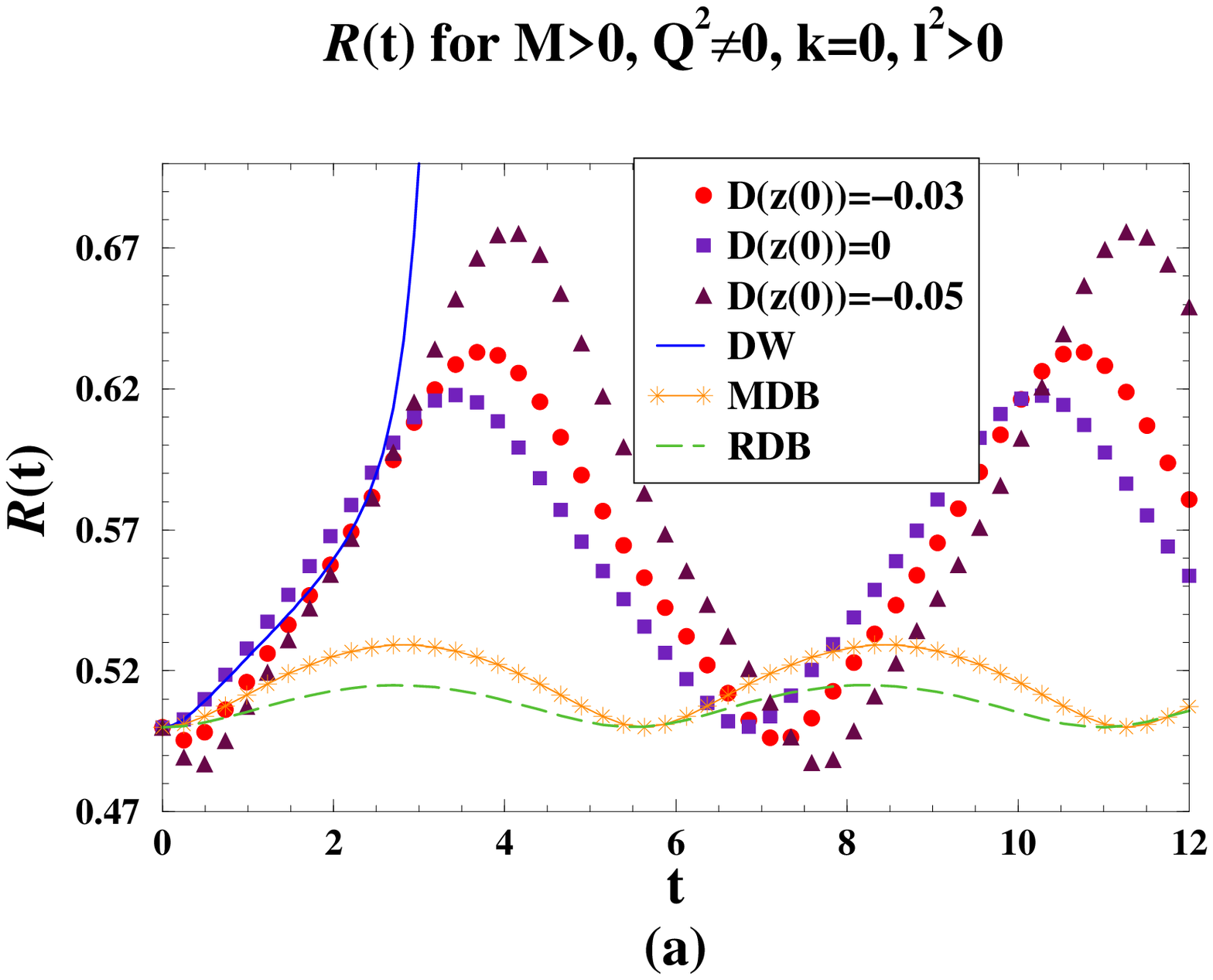}}\nonumber
\rotatebox{0}
{\includegraphics[width=.50\textwidth]{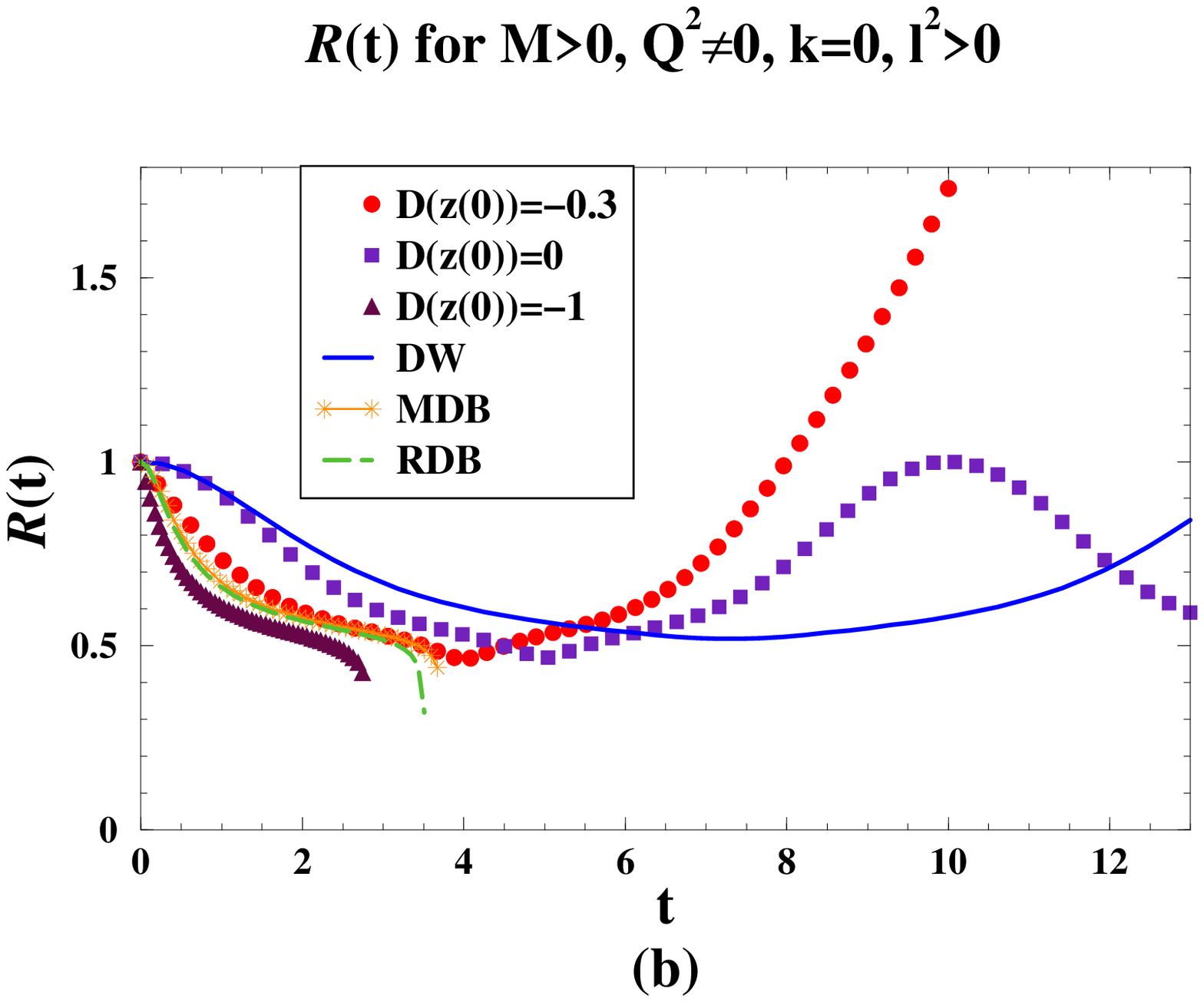}}\nonumber
\end{eqnarray}
\end{center}
\end{figure*}

For $M$ and $Q^2$ non-zero we have solved (\ref{brem}) in the typical
cases of a domain wall ($\omega=-1$), matter ($\omega=0$) and
radiation ($\omega=1/3$) dominated branes. In Fig.\ref{hh} we show
all the possible forms of $h(z)$ due to $M$, $Q^2$, $k$ and $l^2$
combinations, which we study in this section. Some of our results are
also illustrated in Fig. \ref{fig6} together with the solution
of the geodesic 
equation (\ref{geo}) in order to verify the possibility of having
shortcuts. We have also calculated the time delays and the ratio
between graviton and photon horizons for the examples of
shortcuts appearing in Fig. \ref{fig6}, these are shown in Table \ref{td}
together with the graviton bulk flight time and its corresponding
brane time according to the equation (\ref{dtdtau}).

We have classified all cases according to the sign of the $M$ 
parameter and to whether we are in dS or AdS bulks. Moreover, we
studied the zero charge black hole as well as the
Reissner-Nordstr\"om-type solutions, namely 8 cases. In Tables
\ref{noncharged} and \ref{charged} it is displayed the behavior of 
the solutions of the brane equation of motion (\ref{brem}) for a
domain wall, a matter, and a radiation dominated branes. We also
show the geodesic behavior and remark the cases in which shortcuts
are possible. 

\begin{table*}[t]
\caption{Scale factor and geodesics evolution (uncharged case). The
arrow indicates the behavior tendency. There can be more than one
behavior depending on the initial conditions. The star or the dagger
in the last column indicates the possibility of shortcuts for the
matter and radiation dominated branes or the domain wall, respectively.}
\begin{tabular}{ccccccccc}\hline
$M$ &$Q^2$&$k$ &$l^2$ &$h(R)$   &DW&MDB&RDB&Geodesic\\ \hline  
 +  & 0  & 1   & +   &{\footnotesize AdS-Schwarzschild} & $\rightarrow
 r_H$/grow&$\rightarrow 
 r_H$&$\rightarrow r_H$&$\rightarrow {r_H} ^*$/grow\\ 
+&0&0,-1&+&{\footnotesize AdS-topological black hole}& $\rightarrow
 r_H$&$\rightarrow 
 r_H$&$\rightarrow r_H$&$\rightarrow {r_H} ^*$ \\ \hline
 -  & 0  & 0,1 & +   &{\footnotesize AdS-naked singularity}&
 grow&$\rightarrow 0$&$\rightarrow  0$ &grow$^*$/$\rightarrow 0$\\ 
 -  & 0  & -1  & +   &{\footnotesize AdS-topological black hole}& $\rightarrow
 r_H$&$\rightarrow 
 r_H$&$\rightarrow r_H$&$\rightarrow r_H$ \\ \hline
 +  & 0  & 1   & -   &{\footnotesize dS-Schwarzschild}  & $\rightarrow
 r_H$/$\rightarrow 
 r_c$&$\rightarrow r_H$/$\rightarrow r_c$&$\rightarrow
 r_H$/$\rightarrow r_c$ &$\rightarrow r_H$/$\rightarrow {r_c}^{*\dagger}$\\  
 +  & 0  & 0,-1& -   &{\footnotesize dS-cosmological singularity}& {\footnotesize no solution}
 &{\footnotesize no solution}&{\footnotesize no solution}&{\footnotesize no solution}\\ \hline  
 -  & 0  &0,$\pm$1& -&{\footnotesize dS-naked singularity}&
 $\rightarrow r_c$&$\rightarrow 
 0$/$\rightarrow r_c$&$\rightarrow 0$/$\rightarrow r_c$ &$\rightarrow
 0$/$\rightarrow {r_c}^{*\dagger}$ \\
\hline 
\end{tabular} 
\label{noncharged}
\end{table*}
\begin{table*}[t]
\caption{Scale factor and geodesics evolution (charged case).}
\begin{tabular}{ccccccccc}\hline
$M$ &$Q^2$&$k$ &$l^2$ &$h(R)$   &DW&MDB&RDB&Geodesic\\ \hline  
 +& +& {\footnotesize 0}& +&
{\footnotesize AdS-naked singularity}& {\footnotesize $\rightarrow
 \infty$/bounce}& {\footnotesize bounce/$\rightarrow 0$}& {\footnotesize
bounce/$\rightarrow 0$}&{\footnotesize bounce$^{*\dagger}$/}\\
 & & & & & & & & {\footnotesize $\rightarrow\infty
 ^\dagger$/$\rightarrow 0$}\\
 +  & +  &{\footnotesize 1}   & +   &{\footnotesize AdS-naked singularity}& $\rightarrow \infty$&$\rightarrow 0$&$\rightarrow 0$&grow$^*$\\ 
 +  & +  &{\footnotesize -1}  & +   &{\footnotesize AdS-Top.charged
 black hole}& $\rightarrow r_H$&$\rightarrow r_H$&$\rightarrow
 r_H$&$\rightarrow r_H$ \\ 
 +  &$\ll$&{\footnotesize 0,-1}&+    &{\footnotesize AdS-Top.charged
 black hole} & $\rightarrow r_H$&$\rightarrow r_H$&$\rightarrow r_H$
 &$\rightarrow {r_H}^*$ \\ 
 +  &$\ll$&{\footnotesize 1}& +   &{\footnotesize
 AdS-Reissner-Nordstr\"om}& $\rightarrow r_H$/$\rightarrow
 \infty$&$\rightarrow r_H$&$\rightarrow r_H$ &$\rightarrow {r_H}^*$ \\ \hline
 -  & +  &{\footnotesize 0,1} & +   &{\footnotesize AdS-naked
 singularity}& $\rightarrow \infty$&$\rightarrow 0$&$\rightarrow 0$&grow$^*$\\ 
 -  & +  &{\footnotesize -1}& +   &{\footnotesize AdS-Top.charged
 black hole}& $\rightarrow r_H$&$\rightarrow r_H$&$\rightarrow r_H$
 &$\rightarrow {r_H}^*$ \\ \hline
 +  & +  &{\footnotesize 0,-1}&- &{\footnotesize dS-naked
 singularity}&$\rightarrow r_c$&$\rightarrow r_c$&$\rightarrow
 0$/$\rightarrow r_c$ &$\rightarrow {r_c}^{*\dagger}$\\ 
 +& +& {\footnotesize 1}& - &{\footnotesize dS-naked singularity}&
 $\rightarrow r_c$& {\footnotesize bounce/}& $\rightarrow
 0$/$\rightarrow r_c$ &$\rightarrow {r_c}^{*\dagger}$\\
& & & & & & {\footnotesize $\rightarrow 0$/$\rightarrow r_c$} & &\\
+&$\ll$&{\footnotesize 0,-1}&-&{\footnotesize dS-naked
 singularity}&$\rightarrow r_c$&$\rightarrow r_c$&$\rightarrow
 0$/$\rightarrow r_c$ &$\rightarrow {r_c}^{*\dagger}$/$\rightarrow 0$\\ 
 +& $\ll$& {\footnotesize 1}& -&{\footnotesize
 dS-Reissner-Nordstr\"om}& $\rightarrow r_H$/$\rightarrow 
 r_c$& $\rightarrow r_H$/$\rightarrow r_c$& $\rightarrow
 r_H$/$\rightarrow r_c$ &$\rightarrow {r_H}^*$/$\rightarrow
 {r_c}^{*\dagger}$\\ \hline 
 -  & +  &{\footnotesize 0,$\pm$1}& -&{\footnotesize dS-naked
 singularity}&$\rightarrow r_c$&$\rightarrow 
 0$/$\rightarrow r_c$&$\rightarrow 0$/$\rightarrow r_c$ &$\rightarrow
 {r_c}^{*\dagger}$/$\rightarrow 0$ \\ \hline
\end{tabular} 
\label{charged}
\end{table*}
\begin{center}
\begin{table*}[t!]
\caption{\label{td}Bulk time $t$, brane time $\tau$, time delays $\Delta\tau$
and ratio between graviton and photon horizons $g/\gamma$ for shortcut
geodesics.}  
\begin{tabular}{ccccccccccccc}\hline
&\multicolumn{4}{c|}{\bf DW} & \multicolumn{4}{c|}{\bf MDB} &
\multicolumn{4}{c}{\bf RDB}\\ \hline
{Fig.}&$t$&$\tau$&$\Delta\tau$&$g/\gamma$&$t$&$\tau$&$\Delta\tau$&$g/\gamma$&$t$&$\tau$&$\Delta\tau$&$g/\gamma$ \\\hline
{\ref{fig6}a}&{2.03}  &{.464}&{.006}&{1.013} &{.80}
&{.232} &{.002} &{1.008}&{.64} &{.192}  &{.001}&{1.007} \\   
{\ref{fig6}a}&{-}     &{-}&{-}    &{-} &{1.11}&{.318} &{.007} &{1.022}
&{.94}   &{.283}  &{.006}&{1.020} \\  
{\ref{fig6}b}&{5.5}   &{1.6}&{.2}   &{1.206} &{-}&{-}    &{-}    &{-}
&{-}&{-}    &{-}&{-} \\  
{\ref{fig6}b}&{6.2}   &{2.5}&{.1}   &{1.074} &{-}&{-}    &{-}    &{-}
&{-}     &{-}     &{-}&{-}\\ \hline
\end{tabular} 
\end{table*}
\end{center}

\section{Discussion and Conclusions}

In the present work we have studied the behavior of a brane embedded
in a six dimensional de Sitter or Anti de Sitter spacetime containing
a singularity. The system
of equations describing this behavior from the point of view of an
observer in the bulk appears to be highly non-linear. 

By solving the full non-linear system we found different behaviors for
the several scenarios appearing due to all the combinations of $M$,
$Q^2$, $k$ and $l^2$ taken into account. The results show branes
getting away from the singularity, falling into it, converging to
cosmological horizons when they exist or even bouncing between a
minimum and maximum values.
The bouncing behavior is not surprising since in 5 dimensions a similar
behavior has been obtained in recent investigations \cite{chargedbh},
where universes bouncing from a contracting to an expanding phase
without encountering past and/or future singularities appear. In this
way these results could provide support for a singularity-free
cosmology or to the so-called cyclic universe scenarios \cite{st}.

Finally, we also studied the geodesic behavior. Contrarily to the case
of a static brane, where 
shortcuts appeared under very restrictive conditions \cite{accm}, the
present model of a dynamic brane embedded in a static bulk displays
shortcuts in almost all cases and under very mild
conditions. Moreover, despite the fact that the time delay between
graviton and photon flight time is not percentually so big as in other
models \cite{acm} (what is also evident from the ratio between
graviton and photon horizons), it exists and can eventually be measured by the
brane observer, although further considerations are certainly needed in a 
stricter realistic model. On the other hand, the fact that shortcuts are
abundant in the studied setups lends further support to the idea of
solving the horizon problem via thermalization by graviton exchange
\cite{acm,ac}, however, we should stress that this is not a proof of
the solution of the problem yet.

\end{document}